# From Data Completeness to Data Sufficiency: A Task-Driven Imaging Framework for Intraoperative CBCT under Quality-Time-Dose Trade-offs

**Yi Jia[1], Rongjun Ge[2], Yang Chen[1], Yan Xi[3,*] and Wenjun Xia[1,*]**

[1] School of Computer Science and Engineering, Southeast University, Nanjing 210096, China
[2] School of Instrument Science and Engineering, Southeast University, Nanjing 210096, China
[3] Shanghai First-Imaging Information Technology Co., LTD., Shanghai 201201, China
*Author to whom any correspondence should be addressed.

**E-mail:** xiawj@seu.edu.cn

## Abstract

Mobile C-arm cone-beam computed tomography (CBCT) has been widely used for real-time intraoperative 3D imaging. However, current practice often mechanically applies the fan-beam CT criterion of "180° plus fan angle" in pursuit of "data completeness" in reconstruction. This review argues that, under the single circular trajectory of three-dimensional cone-beam geometry, complete data are mathematically unattainable; moreover, blindly increasing sampling may exacerbate the trade-off among intraoperative image quality (Q), imaging time (T), and radiation dose (D). Against this background, this review reframes the evaluation of intraoperative CBCT around "data sufficiency" rather than "data completeness." This perspective moves beyond the excessive pursuit of absolute mathematical and analytic accuracy, and instead emphasizes task-specific minimum image-quality thresholds required for clinical decision-making. By synthesizing evidence from multiple clinical scenarios, this review suggests that approximation errors can be acceptable when clinical decision-making requirements are satisfied, thereby achieving a Q-T-D balance.

## 1. Introduction

### 1.1. Clinical value and application scenarios of intraoperative CBCT

The development of surgical imaging technology is, in essence, a simultaneous pursuit of spatial accuracy and temporal efficiency under clinical constraints. The evolution of surgical guidance from preoperative CT-based planning combined with intraoperative 2D fluoroscopy toward intraoperative real-time 3D imaging, such as mobile C-arm CBCT, has been well documented in previous reviews (Cleary *et al* 2010).

In the traditional surgical workflow, surgeons mainly rely on a combination of preoperative 3D CT-based planning and intraoperative 2D fluoroscopy. High-quality multi-slice CT provides static 3D anatomical information before surgery, whereas intraoperative fluoroscopy offers real-time 2D feedback. However, an inherent information gap remains in this model: preoperative 3D images cannot reflect the real-time intraoperative anatomical state, while intraoperative 2D fluoroscopy suffers from severe tissue superimposition (Weinstein *et al* 1988). In a systematic review of pedicle screw placement, Gelalis *et al* (2012) pointed out that 2D fluoroscopy projects 3D volume onto a 2D plane, where the superimposition will compromise surgeons' spatial perception and make high-precision manipulation difficult. These studies indicate that the traditional preoperative 3D combined with intraoperative 2D workflow cannot provide sufficient and reliable support for complex and dynamic surgical tasks.

Mobile C-arm CBCT brings 3D imaging directly into the crowded operating room, allowing surgeons to access 3D anatomy immediately during surgery. This not only reduces reliance on non-real-time preoperative data, but also compensates for the missing spatial information in conventional fluoroscopy. In the context of spine navigation, mobile C-arm CBCT can be integrated into the intraoperative workflow to provide real-time access to 3D anatomical information (Schafer *et al* 2011). Khoury *et al* (2007) evaluated intraoperative CBCT for image-guided reduction of tibial plateau fracture and showed that, compared with conventional 2D fluoroscopy, intraoperative 3D imaging provides clearer visualization of subtle articular-surface details and 3D displacement of fracture fragments.

From a broader clinical and health-economic perspective, intraoperative CBCT can also improve surgical accuracy, reduce revision rates, and optimize medical resource utilization. A spine surgery strategy based on intraoperative CBCT navigation has been reported to reduce the one-year postoperative revision rate for pedicle screw placement from 6% in the conventional fluoroscopy group to 0.8%, suggesting that intraoperative CBCT not only fills a spatial information gap but also contributes to improved surgical reliability and downstream cost reduction (Dea *et al* 2016). Although intraoperative CBCT and navigation systems involve high acquisition and maintenance costs, their ability to reduce revision surgery caused by screw malposition may generate substantial long-term savings for high-volume centers (Lee *et al* 2020).

Mobile C-arm CBCT has been widely applied across multiple clinical scenarios. In otolaryngology and skull-base surgery, Rafferty *et al* (2006) investigated the navigational potential of intraoperative CBCT in temporal bone surgery and showed that it can provide submillimeter 3D anatomical information at very low radiation doses, offering reliable image guidance in anatomically hazardous regions. In spine and orthopedic surgery, Schafer *et al* (2011) described the integration of mobile C-arm CBCT into spine-navigation workflows for intraoperative assessment of pedicle screws and other fixation devices. Cewe *et al* (2025) further showed that intraoperative CBCT for thoracolumbar spine imaging can approach conventional postoperative multi-detector CT in terms of resolution and contrast-to-noise ratio. In maxillofacial and head-and-neck surgery, Thomas *et al* (2026) systematically reviewed the role of intraoperative CBCT and concluded that it can provide 3D guidance and intraoperative quality assessment. In minimally invasive pulmonary interventions, Setser *et al* (2020) reviewed the use of intraoperative CBCT for visualizing small peripheral pulmonary lesions and guiding bronchoscopic instruments.

### 1.2 Data completeness in CT

In the history of CT technology, the concept of "complete data" was first formalized in 2D image reconstruction. For early 2D parallel-beam CT, the mathematical foundation of analytic reconstruction originated from Cormack's (1963) framework for representing a function by its line integrals. Kak and Slaney (1988) later provided a systematic description of parallel-beam reconstruction theory based on the Radon transform and the central slice theorem. As CT evolved into 2D fan-beam geometry, Parker (1982) derived an optimal short-scan convolution reconstruction algorithm and proposed the "180° plus fan angle" for complete data acquisition. With the further transition to CBCT, the theoretical criterion for data completeness was extended to the Tuy-Smith condition, proposed and refined by Tuy (1983) and Smith (1985) as a sufficient condition for exact 3D cone-beam reconstruction.

Although the Tuy-Smith condition provides a rigorous mathematical criterion for 3D data completeness, engineering practice and some clinical publications on intraoperative CBCT often still adopt the 2D fan-beam standard of "180° plus fan angle" when interpreting short-scan cone-beam acquisitions. Reconstruction under a short-scan circular cone-beam geometry is a typical ill-posed problem, because complete 3D Radon data cannot be acquired, resulting in pronounced cone-beam artifacts and anatomical deformation in noncentral slices (Zellerhoff *et al* 2005). This contradiction shows that mechanically applying the 2D fan-beam $180^{o}$ criterion to 3D cone-beam imaging is mathematically and physically unsound. It also indicated that intraoperative CBCT reconstruction has long been influenced by the pursuit of complete data condition that is unattainable under single circular cone-beam trajectories. Therefore, the evaluation framework for intraoperative CBCT should shift from "data completeness" toward a clinical task-driven strategy of "data sufficiency".

In this review, "data sufficiency" is used as a conceptual lens for re-examining intraoperative CBCT imaging. In this view, the evaluation of intraoperative CBCT should shift from a sole emphasis on mathematically exact reconstruction toward the practical requirements of specific clinical tasks.

Although substantial progress has been made in exact CT reconstruction and mathematical data-completeness solutions, much of this work has not been translated into clinical practice; commercial systems still widely rely on the robust but approximate FDK algorithm (Pan *et al* 2009). This indicates that clinical practice has never blindly pursued data completeness. This view is consistent with a task-based detectability framework, in which image quality is defined by the specific medical task rather than by task-independent physical metrics (Gang *et al* 2014). Physical-parameter measurements detached from specific diagnostic or interventional objectives cannot fully reflect clinical image utility; therefore, evaluation criteria should be linked to clinical tasks rather than to physical sampling conditions alone (Verdun *et al* 2015). Barrett *et al* (2015) systematically discussed the relationship among image quality, radiation dose, and patient risk. Their analysis suggests that, for a specific medical task, image quality should be judged by whether the image enables confident task completion, rather than by whether it is globally perfect. The core idea of the "data sufficiency" paradigm advocated in this review is therefore to move beyond the pursuit of complete data for exact analytic reconstruction and instead use the minimum task-specific image-quality threshold ($Q_{\min}$) required for clinical decision-making as the guiding standard.

## 2. Theoretical Background: Data Completeness in CT

In CT imaging, data completeness determines whether the internal structure of the object can be reconstructed exactly from the projection measurements. The concept of complete data was first rigorously established in 2D reconstruction theory. For early 2D parallel-beam CT, the Radon transform and central slice theorem show that scanning within 180° range is theoretically sufficient for data completeness. The classic review by Brooks and Di Chiro (1976) systematically introduced this theorem into medical radiographic imaging and established that exact analytic CT reconstruction relies on complete semicircular sampling of frequency space.

As CT evolved into fan-beam CT, the minimum angular range for complete data became the "180° plus fan angle" condition, which ensures sufficient coverage of the required line-integral data. Noo *et al* (2002) further mathematically proved that, projection range smaller than 180° plus fan angle produces non-interpolatable data gaps in Radon space, resulting in frequency information loss and severe reconstruction artifacts.

The data completeness of CBCT rests on a more stringent mathematical foundation: the Tuy-Smith condition. Tuy (1983) derived an exact inverse formula for reconstructing a 3D object-density function from CBCT projection data. The formula holds only under specific geometric conditions: every plane intersecting the object must also intersect the X-ray source.

Based on Tuy's 3D inverse formula, Smith (1985) further formulated a geometric and topological sufficiency condition for exact reconstruction. That is, a 3D volume can be uniquely and exactly reconstructed if and only if every 2D plane intersecting the object contains at least one cone-beam X-ray source point.

The Tuy-Smith condition provides a mathematical criterion for data completeness in 3D cone-beam reconstruction. Physically, it means that X-rays source trajectory must sample the object from sufficiently diverse spatial positions so that every object-intersecting plane is represented in the measured data. Together with Grangeat's (1991) mapping between cone-beam projections and Radon space, the Tuy-Smith condition essentially ensures the complete sampling support in the 3D Radon domain. If any object-intersecting planes fail to intersect the source trajectory, unsampled Radon-space gaps remain, making exact analytical reconstruction mathematically impossible.

Under the rigorous requirements of the Tuy-Smith condition, a single circular scan trajectory is inherently incomplete for exact 3D cone-beam reconstruction. Suppose that the circular source trajectory line in the plane z = 0. For any object point away from this plane, there will be a constructed plane passing through that point and parallel to the z = 0 plane. This plane intersects the object but never intersects the circular source trajectory, and therefore violates the Tuy-Smith condition.

Strict satisfaction of the Tuy-Smith condition generally requires a nonplanar source trajectory, whereas any single planar trajectory inevitably leaves non-interpolatable data gaps in Radon space (Kudo and Saito 1990). Such incomplete coverage can lead to missing high-frequency information along the axial direction in 3D frequency space, forming a "missing dimension" during reconstruction (Bartolac *et al* 2009). Wu *et al* (2023) further quantified the degree of data incompleteness for single circular trajectories compared with more complex trajectories. These studies show that, regardless of whether the angular span of a single circular scan is 180° or 360°, and regardless of sampling density or radiation dose, a single circular source trajectory cannot achieve complete data for exact 3D cone-beam reconstruction.

Although a single circular trajectory cannot provide 3D complete data satisfying the Tuy-Smith condition, Feldkamp *et al* (1984) proposed the FDK algorithm for cone-beam reconstruction from circular trajectories. Benefiting from its strong engineering practicality, FDK has long served as a cornerstone of CBCT reconstruction. Before FDK, mature exact reconstruction algorithms such as filtered backprojection (FBP) had already been established for 2D parallel-beam and fan-beam CT. FBP originated from the spatial-domain one-dimensional convolution backprojection framework proposed by Ramachandran and Lakshminarayanan (1971). Shepp and Logan (1974) subsequently introduced the Shepp-Logan filter to reduce the high-frequency noise, enabling FBP to achieve successful clinical performance in 2D CT under complete-data conditions. Building on the theoretical foundation, Feldkamp *et al* (1984) extended the filtered backprojection algorithm to circular 3D cone-beam geometry.

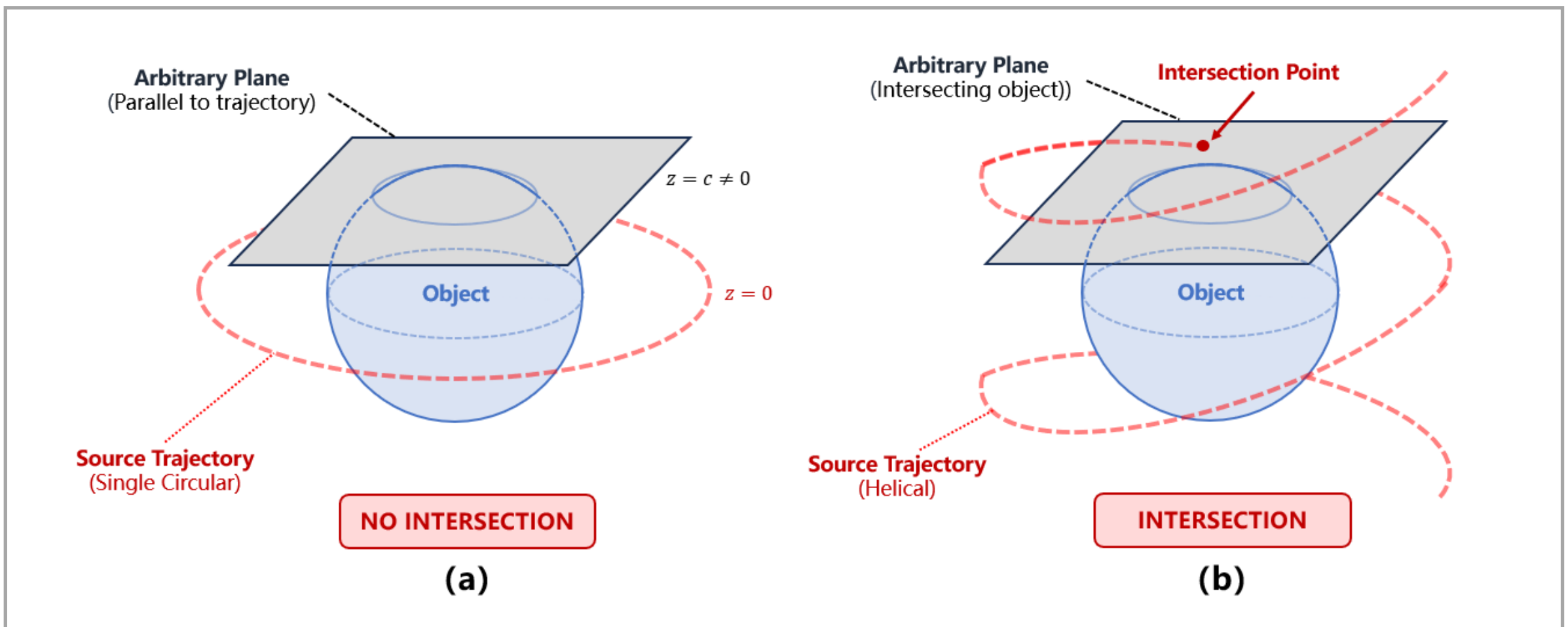

**Figure 1.** Comparison of data completeness under different cone-beam CT source trajectories. (a) A single circular trajectory fails to satisfy the Tuy-Smith condition: an arbitrary plane parallel to the source trajectory can intersect the object without intersecting the source trajectory, leading to incomplete cone-beam data. (b) A helical trajectory can satisfy the Tuy-Smith condition: an arbitrary plane intersecting the object also intersects the source trajectory at one or more points, thereby providing the geometric basis for complete 3D cone-beam reconstruction.

Feldkamp *et al* (1984) acknowledged that FDK is an approximate reconstruction algorithm for circular cone-beam geometry. On noncentral planes, the algorithm applies backprojection approximation that treats oblique rays in a manner similar to the central plane, thereby introducing FDK approximation error. This error essentially arises from data incompleteness on noncentral planes.

The approximate nature of FDK may introduce noticeable errors in clinical images. FDK can amplify artifacts in cone-angle regions, where off-plane structures may appear as cupping-like or streak-like artifacts (Schulze *et al* 2011). In scenarios involving metal instruments or dense bone, Grass *et al* (1999) demonstrated that FDK can cause nonlinear spreading and intensity reduction of high-contrast structural edges along the axial direction. These observations verify that, under the framework of the single circular trajectory and approximate reconstruction, strict data completeness cannot be achieved in intraoperative CBCT.

## 3. Clinical Constraints in Intraoperative CBCT

### 3.1 Various CBCT Application Scenarios

CBCT has been widely used in multiple clinical fields, including image-guided radiotherapy CBCT (IGRT-CBCT), dental CBCT, and intraoperative CBCT. Because the clinical objectives differ across these applications, the anatomical focus, imaging quality, and radiation-dose control requirements of different systems vary substantially, leading to marked differences in data-sufficiency.

In radiotherapy, the core tasks of IGRT-CBCT are high-precision target localization and dose planning for adaptive radiotherapy (ART). Its data-sufficiency requirements mainly involve grey-level fidelity and temporal consistency. Stock *et al* (2009) validated the image quality and stability of IGRT devices for establishing accurate patient-position mapping in image-guided radiotherapy. If CBCT images are to be used directly for accurate target dose calculation, the high-intensity cone-beam scatter associated with a large field of view (FOV) must be overcome (Bissonnette *et al* 2012). Sonke *et al* (2005) showed that respiratory phase sorting along the time axis retains only sparse effective projection samples within each phase, thereby producing severe streak artifacts in reconstructed images.

Dental CBCT primarily aims to reproduce fine hard-tissue anatomy accurately while remaining constrained by strict radiation-protection principles. In dental implant applications, Unger *et al* (2023) showed that CBCT can support high-precision CAD/CAM surgical-guide design and implant-position verification. Scarfe *et al* (2006) indicated that, to visualize subtle hard-tissue features such as dental trauma and accessory canals, dental CBCT system must provide very high spatial resolution, often requiring voxel sizes below 100 μm. Pauwels *et al* (2012), in their analysis of radiation dose and image quality in oral and maxillofacial CBCT, noted that, under the ALARA principle, dental examinations widely use variable small-FOV or region-of-interest (ROI) scanning strategies, which introduces spatial data incompleteness.

The primary objective of intraoperative CBCT is to provide real-time and accurate spatial verification of hard-tissue anatomy and high-density implants within the constrained operating room. Compared with radiotherapy and dental CBCT, which have relatively standardized scanning trajectories, intraoperative CBCT meets greater data-sufficiency challenges. The special constraints of the intraoperative setting can lead to serious data insufficiency in mobile C-arm CBCT-guided spine surgery (Schafer *et al* 2011). In interventional environments, intraoperative CBCT is often forced to use limited angular sampling, resulting in elongation distortion and strong anisotropic blurring (Siewerdsen *et al* 2013).

### 3.2 Special Characteristics of the Intraoperative Setting

**Table 1.** This table comprehensively compares the significant differences across three typical application scenarios: IGRT-CBCT, Dental-CBCT, and Intraoperative CBCT.

| Comparison Dimension | IGRT-CBCT | Dental-CBCT | Intraoperative CBCT |
|---|---|---|---|
| Time Constraint (T) | Moderate; minute-level acquisition, often with respiratory gating | Moderate; static scanning is generally acceptable | Strict; constrained by surgical workflow, physiological motion, and sterile-field requirements |
| Dose Constraint (D) | Moderate; imaging dose is relatively small compared with therapeutic dose | Strict; governed by ALARA/ALADA principles and limited field of view | Strict; patient and staff exposure must both be considered |
| Trajectory Characteristics | Standard 360°/180° circular trajectory | Standard circular scanning | Highly Restricted, often limited to small-angle or incomplete circular trajectories |

Unlike the static and highly controllable environment of a diagnostic radiology suite, the operating room is dynamic, crowded, and less controllable. Intraoperative CBCT commonly uses mobile C-arm systems. Schafer *et al* (2011) noted that, during interventional and surgical procedures, mobile C-arms must strictly respect sterile-field isolation. Hatamikia *et al* (2021) also noted that the rotation trajectory of a mobile C-arm is often strongly constrained by operating-table base, complex cables and tubes, or the patient's limbs. These safety and workflow constraints directly prevent mobile C-arms from executing complex scan trajectories in intraoperative settings, forcing them to adopt limited-angle circular trajectories and making ideal high-quality reconstruction more difficult.

Intraoperative CBCT also increases operative time. Intraoperative CBCT scanning and image registration typically add 1 to 20 minutes to the procedure (Thomas *et al* 2026). Blindly pursuing data completeness may prolong each scan, increase motion artifacts, and raise surgical risk. Longer scan durations increase motion artifacts caused by involuntary patient micromovements, cardiac pulsation, and respiration, thereby degrading image quality (Spin-Neto and Wenzel 2016). Cheng *et al* (2018), based on a large clinical data analysis, confirmed that unnecessary prolongation of operative and anesthesia time is a risk factor for postoperative wound infection and other deep complications. Therefore, in the operating room, extending scan time in pursuit of data completeness not only weakens image quality but also increases overall surgical risk.

Radiation protection in conventional diagnostic CT mainly concerns the patient. In the intraoperative setting, however, surgeons, nurses and other staff may remain in the operating room or behind radiation shields, exposing both patients and medical personnel to radiation risk. The International Commission on Radiological Protection (2007), in Publication 103, established the ALARA (as low as reasonably achievable) principle. Jaju and Jaju (2015) further argued that, in complex clinical environments, imaging optimization should move toward the ALADA (as low as diagnostically acceptable) principle. Intraoperative imaging systems can generate non-negligible scatter radiation; thus, dose control must conform to strict ALARA principles, and image quality cannot be improved simply by substantially increasing dose (Daly *et al* 2006). Cewe *et al* (2022) evaluated cumulative occupational radiation exposure in intraoperative CBCT workflows and showed that, without dose control and workflow optimization, lead surgeons and nurses may face an increased risk of approaching or exceeding annual occupational exposure limits.

## 4. The Sufficiency Framework

### 4.1 Paradigm Shift from Data Completeness to Data Sufficiency

The field of CBCT reconstruction has long focused on the traditional strategy of "data completeness" in pursuit of mathematically exact reconstruction. In their systematic review, Defrise and Gullberg (2006) noted that much work in CBCT reconstruction has focused on designing complex scan trajectories that strictly satisfy the Tuy-Smith condition and on deriving exact analytic inverse formulas in 3D Radon space. Among these efforts, Katsevich (2002b) derived an exact filtered-backprojection inversion formula for helical cone-beam CT, representing an important milestone in mathematically exact cone-beam reconstruction.

However, such data completeness-oriented methods are difficult to reconcile with the practical constraints of the intraoperative imaging. For example, Katsevich's (2002a) algorithm relies on helical scanning, whereas such nonplanar trajectories are difficult to realize in a crowded operating room. In addition, Katsevich (2002a) noted that a key preprocessing step in the reconstruction formula involves differentiation of projection data along the trajectory, which can increase sensitivity to noise and computational cost. This is particularly challenging for intraoperative CBCT, where low-dose acquisition leads to unavoidable quantum noise, while the surgical workflow imposes strict requirements on computational efficiency and reconstruction speed.

From the perspective of clinical implementation, Pan *et al* (2009) noted that although substantial progress has been made in exact CT reconstruction, most of these achievements have not been translated into routine clinical practice; commercial CT and intraoperative mobile C-arm systems still widely rely on the theoretically approximate FDK algorithm. This gap reflects a fundamental mismatch between mathematically complete reconstruction and the physical, temporal, and dose constraints of real clinical systems. FDK has been widely adopted not because it satisfies ideal data-completeness requirements, but because it provides a robust and computationally efficient approximation under practical conditions. Therefore, CBCT reconstruction, especially in the intraoperative setting, should be evaluated not solely by mathematical completeness, but by whether the acquired data and reconstructed images are sufficient for the intended clinical task.

### 4.2 Operational Definition of Data Sufficiency

Task-based image-quality assessment provides an important basis for defining data sufficiency in terms of clinical task performance rather than mathematical completeness (Gang *et al* 2014). They emphasized that the "quality" of a medical image is not an independent mathematical metric but is defined by the specific medical task in which the image is used. The data sufficiency strategy proposed in this review is guided by this concept of task-defined quality. Its core idea is to move beyond the sole pursuit of mathematically exact reconstruction and instead aim for clinically meaningful optimality. In this framework, controlled approximation errors at the algorithmic level can be accepted as long as an image satisfies a set of task-driven quality requirements needed for clinical decision-making.

Accordingly, this review adopts the following operational definition of "data sufficiency": within a radiation-dose budget (D) consistent with the ALARA principle and a clinically acceptable imaging time window (T), a set of projection data($P_\beta$) can be considered sufficient if and only if the reconstructed 3D image ($f(x, y, z)$) satisfies the set of minimum image-quality thresholds ($Q_{\min}$) required for a specific clinical task.

### 4.3 Establishment of the Metric System $Q_{\min}$

The task-based detectability framework established by Gang *et al* (2014) provides a mathematical basis for defining task-specific quality thresholds. Gang *et al* (2015) further argued that global high resolution should not be pursued blindly; instead, tube current, orbital angulation, and reconstruction parameters should be concentrated in the regions most important for a given detection task. Oenning *et al* (2021) proposed the ALADAIP concept ("as low as diagnostically acceptable, indication-oriented and patient-specific") for pediatric CT on the basis of ALARA.

Under the data-sufficiency strategy, image-quality assessment shifts toward a set of minimum metric thresholds required for a specific clinical task, denoted as $Q_{\min}$. This threshold is not a single static index but a multidimensional and task-dependent set of criteria. Its core components include spatial resolution, low-contrast resolution, image uniformity, geometric accuracy, and artifact-control level, with different clinical tasks requiring different combinations and levels of these metrics.

### 4.4 Significance of the Paradigm Shift

The evolution from mathematically rigorous "data completeness" to clinically oriented "data sufficiency" is not a compromise imposed by the limitations of mobile C-arms in the intraoperative setting. Rather, it represents an evolution of the evaluation framework for intraoperative CBCT. This strategy shift has important theoretical and engineering implications.

First, the data-sufficiency strategy establishes task-oriented evaluation criteria. As reflected in the task-driven imaging concept proposed by Gang *et al* (2015), projection data should no longer be evaluated solely according to whether they satisfy the Tuy-Smith condition, but according to whether they provide sufficient information for a specific clinical task. Second, the sufficiency threshold can evolve with advances in reconstruction technology. Deep-learning-based reconstruction models can learn and recover high-fidelity structures from extremely sparse or truncated projections (Wang *et al* 2018). With the development of AI algorithms, the amount of physical front-end scanning required for a given task may be further reduced. This demonstrates that the data-sufficiency strategy is a dynamic framework that integrates physical acquisition requirements with computational imaging capabilities.

## 5. Case Studies and Validation

### 5.1 Data-Sufficiency Analysis in Spine and Orthopedic Trauma Surgery

Spine and orthopedic trauma surgery represent classic application scenarios for intraoperative CBCT, mainly including pedicle screw placement and reduction of complex comminuted pelvic or calcaneal fractures. In this setting, the clinical task of intraoperative CBCT is to assess hard-tissue integrity and the spatial position of fixation devices intraoperatively. Specifically, it is used to verify whether pedicle screws are accurately placed and fully contained within cortical bone, and whether small fragments in complex 3D comminuted fractures of the pelvis, calcaneus, and related structures have been reduced. These tasks require high geometric accuracy, artifact control, and spatial resolution, while placing relatively lower demands on soft-tissue low-contrast resolution and global image uniformity. Although computation time and radiation dose must still be carefully controlled, these applications may justify a higher image-quality requirement because spatial accuracy is directly related to surgical safety.

To satisfy the navigational needs of spine surgery, mobile C-arm CBCT must provide submillimeter spatial resolution, with pixel-level precision approaching approximately 0.3 × 0.3 mm (Schafer *et al* 2011). Wang *et al* (2014) showed that, because of X-ray scatter, large cone-angle truncation, and electronic noise, intraoperative CBCT has inherent limitations in global image uniformity and soft-tissue low-contrast resolution. Although intraoperative CBCT has higher image noise and lower global contrast-to-noise ratio (CNR), its excellent spatial resolution can better depict the key anatomical details required clinically in spine surgery (Casiraghi *et al* 2021). Thus, under the data-sufficiency strategy, resources for hard-tissue spine tasks should be directed toward high-frequency spatial resolution, whereas low-frequency image noise can be tolerated to some extent.

**Table 2.** Because different surgical scenarios involve distinct physical properties of target tissues and different interventional approaches, their multidimensional image-quality requirements exhibit strong task specificity.

| intraoperative application scenario | Spatial resolution | Low-contrast resolution (soft tissue) | Image uniformity (quantitative) | Geometric accuracy (navigation/registration) |
|---|---|---|---|---|
| Spine and orthopedic trauma | High | Low | Low | Very high |
| Neurosurgery and otolaryngology | Very high | Moderate | Low | Very high |
| Maxillofacial severe trauma reconstruction | High | Low | Medium-low | Moderate |
| Thoracoabdominal soft tissue (hybrid OR) | Moderate | Very high | High | Moderate |
| Interventional navigation and vascular surgery | Medium | Very high | Medium | High |

For intraoperative percutaneous pedicle screw placement, blooming around screw shafts and streak artifacts at screw tips can seriously hinder clinicians' assessment of cortical breach (Charles *et al* 2022). This finding demonstrates the importance of artifact control for the reliability of intraoperative CBCT in spine surgery.

Mendelsohn *et al* (2016) showed that intraoperative CBCT improves screw-placement accuracy but significantly increases overall radiation dose, making dose constraints particularly important under data-sufficiency criteria. Vorbau *et al* (2024) showed that in spine surgery, intraoperative CBCT requires a higher dose trade-off than conventional CT when handling high-frequency imaging tasks. For high-contrast tasks such as hard-tissue mechanical fixation, the system may prioritize resources for geometric accuracy, while accepting reduced detectability for low-frequency soft-tissue tasks.

**5.2 Data-Sufficiency Analysis in Neurosurgery and Minimally Invasive Otolaryngologic Surgery**

In neurosurgery and minimally invasive otolaryngologic surgery, the main tasks of intraoperative CBCT include guiding precise deep brain stimulation (DBS) microelectrode implantation, reconstructing microscopic ossicular-chain anatomy, and verifying bony boundaries in real time during endonasal transsphenoidal skull-base tumor resection. Because the relevant anatomical structures are small, these scenarios require high spatial resolution. Because functional neurosurgical procedures such as DBS rely strongly on stereotactic systems, they also require very high 3D geometric mapping accuracy. Clinically, the scanning region may involve radiosensitive organs such as the lens and thyroid, imposing highly stringent ultra-low-dose constraints. Reconstruction time must also be sufficiently short to capture intraoperative anatomical changes such as brain shift induced by cerebrospinal-fluid loss.

For brain parenchymal imaging in neurosurgery, Sharma and Deogaonkar (2016) showed that intraoperative CBCT provides 3D targeting accuracy comparable to standard fixed multislice CT. This demonstrates that, even when intraoperative scanning is limited by angle and trajectory and does not provide mathematically complete data, the resulting image quality can still meet the sufficiency threshold for localizing small electrodes in high-precision neurosurgical geometric mapping. Zhang *et al* (2023) showed that, due to X-ray scatter, large cone-angle beam hardening, and related effects, intraoperative CBCT has inherent physical limitations in low-contrast resolution for brain soft tissue and is difficult to use independently for clear diagnosis of brain-parenchymal boundaries or small intracranial hemorrhage. Nevertheless, intraoperative CBCT with metal-artifact reduction can not only verify electrode position with high 3D geometric accuracy, but also rapidly capture transient brain shift caused by cerebrospinal-fluid loss or intracranial air during surgery (Kawasaki *et al* 2022). This indicates that, for the specific DBS task, $Q_{\min}$ is centered on metal-artifact control and high-contrast 3D electrode localization, while reduced soft-tissue CNR can be accepted.

In otolaryngologic and skull-base surgery, Lee *et al* (2012) showed that intraoperative CBCT can provide surgeons with low-dose, high-resolution real-time images during complex endonasal transsphenoidal skull-base tumor resection. Kemp *et al* (2020) showed that, for small anatomical structures such as the ossicular chain, high-resolution CBCT can provide isotropic spatial resolution superior to conventional multislice CT, satisfying the basic requirements for bony microanatomical visualization in otologic minimally invasive surgery and implant verification. Muhanna *et al* (2021) reported that, although CBCT performs poorly in soft-tissue visualization, it performs well in displaying complex bony skull-base boundaries. This shows that clinical utility in this scenario does not require comprehensive data completeness; rather, data need only satisfy local high-resolution requirements defined by $Q_{\min}$.

Mihailidis *et al* (2024) noted that, in CBCT imaging involving the inner ear and skull base, radiation exposure to highly sensitive organs such as the lens and thyroid must be carefully considered. This further shows that blindly pursuing a complete-data standard is not feasible in this scenario.

### 5.3 Data-Sufficiency Analysis in Maxillofacial and Severe Head-and-Neck Trauma Reconstruction

In maxillofacial and severe head-and-neck trauma reconstruction, the clinical tasks of intraoperative CBCT focus on 3D symmetric restoration of complex multifocal facial fractures, precise osteotomy alignment in orthognathic surgery, and morphological verification of reconstruction using free fibular flaps combined with large titanium plates. These tasks require clear visualization of whether extremely thin orbital-floor bone fragments have been reduced and whether multiple facial sutures and micro-titanium screws are accurately aligned, thereby demanding high spatial resolution and geometric accuracy. Metal-induced streak artifacts are also common in this scenario, creating a high demand for artifact control. Because these complex reconstructions are usually long procedures performed under general anesthesia, they may allow a slightly larger time budget for high-precision physical scanning and more advanced reconstruction algorithms before wound closure, although cumulative radiation dose from repeated local scans must still be controlled.

Daly *et al* (2006) described the fundamental trade-off between radiation dose and image-quality metrics. They showed that, under very low-dose incomplete-scan conditions, CBCT can still achieve submillimeter 3D spatial resolution and clearly depict bony anatomical details, whereas reliable soft-tissue CNR would require radiation dose to increase substantially. Pohlenz *et al* (2008), in a study of intraoperative imaging for major mandibular surgery, showed that intraoperative CBCT can provide high-spatial-resolution images sufficient for clinical needs and can be used directly during surgery to assess the 3D alignment of thin bony plates.

Van Hout *et al* (2022), in a study of zygomaticomaxillary complex (ZMC) fracture repair, claimed that despite limited soft-tissue contrast, intraoperative CBCT provides sufficient high-frequency spatial information. This demonstrates that, for maxillofacial fracture reconstruction, clinical data sufficiency is primarily determined by whether the system can meet the spatial-resolution requirement of key local bony anatomy, rather than by whether the image is globally complete. Haines *et al* (2024), in a narrative review of intraoperative CT management of complex facial fractures, noted that intraoperative CBCT image quality must support fine-slice 3D reconstruction with slice thickness below 1 mm; only with this level of image quality can surgeons detect submillimeter anatomical asymmetry or poor titanium-plate adaptation before closure.

Cuddy *et al* (2021) further clarified the boundary effect of metal artifacts on the utility of intraoperative CBCT. They reported that even when obvious radial streaks and dark bands appear in the image, if artifact control at key bony buttresses is sufficient to distinguish hardware edges, surgeons can confirm bone-fragment reduction and the final hardware position. This indicates that, in head-and-neck reconstruction with extensive metal implants, $Q_{\min}$ is not centered on mathematically complete global artifact removal, but on whether residual artifacts interfere with clinically critical structures and hardware boundaries.

Thomas *et al* (2026) quantitatively summarized the additional operative time associated with intraoperative CBCT in maxillofacial trauma surgery and its direct impact on restoring normal bony symmetry. Pohanka *et al* (2025) showed that, for repeated scans involving the neck and thyroid, intraoperative CBCT can satisfy the need for 3D bony imaging while keeping cumulative radiation dose far below that of conventional multislice CT. Pietzka *et al* (2024), in an experimental radiation-dose comparison of different 3D imaging devices in the facial skull region, showed that systems used for intraoperative verification in maxillofacial trauma must satisfy clinical diagnostic image quality while still respecting dose constraints.

### 5.4 Sufficiency Analysis in Hybrid Operating-Room Thoracoabdominal Soft-Tissue Surgery

In hybrid operating-room thoracoabdominal soft-tissue surgery, the clinical intervention targets of intraoperative CBCT are mainly low-density soft tissues. Typical tasks include minimally invasive localization and resection of pulmonary ground-glass opacities (GGOs) during thoracoscopy, as well as radiofrequency or microwave ablation of liver parenchymal tumors. The core image-quality requirements in this scenario are low-contrast resolution and low-frequency image uniformity. At the same time, thoracoabdominal surgery is constrained by imaging time and radiation dose. Because cardiopulmonary physiological motion can cause motion blurring, scanning must be completed rapidly within a single breath-hold while maintaining dose control.

In thoracic surgery for small pulmonary-nodule localization, Yang *et al* (2016) and Hsieh *et al* (2017) showed that CBCT in a hybrid operating room can integrate pulmonary-nodule localization, dye marking, or other marking procedures with video-assisted thoracoscopic surgery (VATS) resection into a single-stage workflow, particularly for GGOs that are difficult to palpate intraoperatively and have low density and indistinct margins. Hisano *et al* (2017) noted that GGO lesions have very low CT attenuation, typically ranging from -800 to -200 HU, requiring extremely high soft-tissue low-contrast resolution to ensure that the CNR reaches a clinically effective threshold. In addition, cardiopulmonary motion can readily produce motion blur and obscure subtle lesion boundaries, so CBCT scanning must be completed rapidly within a single breath-hold. These findings also demonstrate that natural anatomical contrast can partially compensate for data insufficiency: even under respiratory constraints and incomplete scanning, the system does not need extremely high spatial resolution as long as the reconstructed image meets the task-specific $Q_{min}$.

For intraoperative imaging of thoracoabdominal solid organs such as the liver, Floridi *et al* (2014) noted that in abdominal minimally invasive procedures such as radiofrequency ablation of liver tumors, the density difference between tumor margins and normal liver parenchyma is usually extremely small and often less than 30 HU. Bapst *et al* (2016) further showed that, in interventional oncology of the liver, CBCT must be combined with different contrast-injection protocols to highlight tumors, hepatic vessels, and liver parenchyma. This indicates that image-quality thresholds for liver soft-tissue tasks depend strongly on contrast phase, uniform parenchymal enhancement, and the visibility of low-contrast lesions. For thoracic soft-tissue surgery, the "sufficiency" of image quality is not equivalent to the global grey-level accuracy of diagnostic CT, but rather depends on whether lesion localization and intraoperative decision-making can be supported within limited time and dose constraints (Kaiho *et al* 2022).

#### 5.5 Data-Sufficiency Analysis in Interventional Navigation and Vascular Surgery

In pulmonary and vascular interventional navigation, the core task of intraoperative CBCT is to provide timely anatomical updates and accurate guidance for surgical instruments such as catheters and puncture needles. Because cardiac pulsation and respiration can affect both anatomy and instrument position, this scenario is subject to strict time-window. As a result, compromises in data completeness are often required under the Q-T-D constraint.

In interventional navigation, spatial resolution does not need to reach the level required for dental implants (25 lp/cm). Instead, the minimum standard is mainly determined by the sizes of the instrument and the lesion. With appropriate optimization of dose and projection number, intraoperative CBCT reconstructions can clearly depict peripheral pulmonary lesions smaller than 1 cm and distinguish solid nodules from pure ground-glass opacities under low-contrast conditions (Setser *et al* 2020). Pritchett *et al* (2018) showed that, after intraoperative CBCT is fused with augmented fluoroscopy, 3D CBCT information can be overlaid onto real-time fluoroscopic images, improving lesion localization and instrument confirmation. This indicates that, in interventional navigation, $Q_{\min}$ is not defined by global diagnostic-grade image quality, but by whether the spatial relationship among lesion boundary, instrument tip, and local anatomical pathway can be stably identified.

In vascular surgery, intraoperative CBCT must also carefully balance image quality and radiation dose. Toroi *et al* (2023) explicitly noted that image-quality requirements for intraoperative CBCT should first be defined and then used to optimize exposure settings, thereby avoiding unnecessary radiation burden while obtaining additional 3D information. Markelj *et al* (2012) also noted that grey-level- or gradient-based registration algorithms are highly sensitive to the consistency of structural gradients within images, while tolerating relatively large errors in absolute HU values. Therefore, in interventional navigation and vascular surgery, data sufficiency should be judged by whether the image can support reliable target localization, instrument guidance, and image registration. As long as artifacts, noise, and nonuniformity do not compromise these key clinical judgments, an image that does not satisfy global diagnostic CT quality can still be sufficient for the intraoperative task.

## 6. Discussion

#### 6.1 New Directions in Algorithm Research: AI Reconstruction and Data Compression

Under the concept of data sufficiency, limited-angle or sparse-angle scanning may become a more clinically appropriate strategy for intraoperative CBCT when guided by specific task requirement. This requires CBCT reconstruction research to shift from the sole pursuit of exact analytic reconstruction toward approximate compensation and prior incorporation under incomplete data. In this context, AI-based reconstruction shows great potential for mitigating the image degradation caused by data incompleteness.

Existing studies indicate that deep-learning models can compensate for streak artifacts and low-contrast degradation produced by conventional FDK under sparse-view or low-dose conditions. For example, Lahiri *et al* (2023) proposed a sparse-view CBCT reconstruction algorithm based on data consistency and generative adversarial networks (GANs). Under severely limited projection views, the method uses deep learning to reduce streak-artifact while enforcing consistency with projection data, enabling the reconstructed images to meet the $Q_{\min}$ requirements of specific surgical tasks. Zhang *et al* (2023) combined physics-based priors with a deep-learning-synthesized DL-Recon 3D reconstruction framework, automatically introducing deep-learning-generated soft-tissue features in regions where incomplete data cause severe FDK artifacts and high uncertainty, thus compensating for insufficient measured data. Chen *et al* (2018) proposed the LEARN network, which unfolds a traditional iterative reconstruction model into a data-driven convolutional neural network, demonstrating that sparse low-dose projection views can still be used to reduce streak artifacts and recover soft-tissue features in the target region.

**6.2 Recommendations for Updating Industry Standards**

Current image-quality evaluation of medical devices still largely follows static physical-phantom testing logic and overemphasizes absolute reconstruction accuracy. To accommodate the paradigm of task-driven imaging and data sufficiency, differentiated scanning protocols and quality-evaluation standards should be developed for intraoperative CBCT. Recent developments in industry standards have already shown a trend toward task-specific image-quality assessment. For example, the AAPM TG-233 report formally advocated replacing absolute physical-metric measurements with a task-based evaluation framework and introduced task transfer function (TTF) and detectability index d' (Samei *et al* 2019). The AAPM TG-238 report further emphasized the fundamental differences between intraoperative C-arm systems and conventional multidetector CT in terms of limited projection angles, complex mechanical deformation, and high-scatter environments(Supanich *et al* 2023). Future standards should avoid universal quality requirements for all intraoperative CBCT applications. Instead, they should establish different $Q_{\min}$ evaluation systems for specific clinical workflows and encourage manufacturers to develop highly customized and efficient task-oriented scanning protocols.

**6.3 Limitations and Future Work**

Although the Q-T-D triangular constraint model discussed in this review offers a theoretical framework for optimizing intraoperative CBCT under the concept of data sufficiency, several limitations remain. First, the quantitative boundary between "sufficient" and "insufficient" data remains unclear. The theory of how incomplete data can become before task failure occurs is still underdeveloped. Future studies need to quantitatively analyze how missing projections, noise, artifacts, and motion blur affect actual intraoperative decision-making across different clinical tasks.

Future work should also develop an intraoperative CBCT evaluation system oriented toward the medical metaverse and surgical digital twins, and construct metaverse-based simulation and validation platforms for intraoperative CBCT. Such platforms could systematically test different scanning protocols, reconstruction algorithms, and navigation strategies in virtual surgical environments. Shao *et al* (2023) noted that the medical metaverse depends on the integration of artificial intelligence, virtual reality, augmented reality, digital twins, and network communication. Hulsen (2024) also pointed out that the medical metaverse can be applied to medical education, virtual consultation, remote collaboration, and personalized medicine, while its clinical deployment still requires solutions to data security, interoperability, and reliability problems. This suggests that, for intraoperative CBCT to be integrated into future digital surgical workflows, it must simultaneously satisfy requirements related to image quality, computational latency, data interoperability, privacy protection, and clinical accountability.

## 7. Conclusion

This review analyzes the complete-data contradiction of mobile C-arm CBCT in intraoperative imaging and points out the theoretical misconception of mechanically applying 2D CT complete-data standards to 3D cone-beam geometry. Because the intraoperative environment is crowded, operative time is limited, and radiation dose is strictly controlled, pursuing complete data satisfying the Tuy-Smith condition is not only infeasible under a single circular scan trajectory, but also incompatible with the Q-T-D trade-off. Therefore, this review supports a task-driven evaluation perspective in which "data sufficiency" becomes more clinically relevant than "data completeness" in intraoperative CBCT. Under this paradigm, reconstructed images should not be judged primarily by whether they enable mathematically exact analytic reconstruction, but by whether they satisfy the task-specific image-quality thresholds ($Q_{\min}$). As long as an image reaches the clinical lower bound required for a given task, controlled reconstruction errors can be accepted, enabling balanced optimization of scan time and radiation dose. This framework provides a theoretical basis for task-oriented scanning protocols, AI-assisted reconstruction, and future standardization of intraoperative CBCT systems.

## References

[1] Bapst B, Lagadec M, Breguet R, Vilgrain V and Ronot M 2016 Cone Beam Computed Tomography (CBCT) in the Field of Interventional Oncology of the Liver *Cardiovasc. Intervent. Radiol.* **39** 8–20

[2] Barrett H H, Myers K J, Hoeschen C, Kupinski M A and Little M P 2015 Task-based measures of image quality and their relation to radiation dose and patient risk *Phys. Med. Biol.* **60** R1–75

[3] Bartolac S, Clackdoyle R, Noo F, Siewerdsen J, Moseley D and Jaffray D 2009 A local shift-variant Fourier model and experimental validation of circular cone-beam computed tomography artifacts *Med. Phys.* **36** 500–12

[4] Bissonnette J-P, Balter P A, Dong L, Langen K M, Lovelock D M, Miften M, Moseley D J, Pouliot J, Sonke J-J and Yoo S 2012 Quality assurance for image-guided radiation therapy utilizing CT-based technologies: A report of the AAPM TG-179 *Med. Phys.* **39** 1946–63

[5] Brooks R A and Chiro G D 1976 Principles of computer assisted tomography (CAT) in radiographic and radioisotopic imaging *Phys. Med. Biol.* **21** 689–732

[6] Casiraghi M, Scarone P, Bellesi L, Piliero M A, Pupillo F, Gaudino D, Fumagalli G, Del Grande F and Presilla S 2021 Effective dose and image quality for intraoperative imaging with a cone-beam CT and a mobile multi-slice CT in spinal surgery: A phantom study *Phys. Med.* **81** 9–19

[7] Cewe P, Skorpil M, Fletcher-Sandersjöö A, El-Hajj V G, Grane P, Fagerlund M, Kaijser M, Elmi-Terander A and Edström E 2025 Image quality assessment in spine surgery: a comparison of intraoperative CBCT and postoperative MDCT *Acta Neurochir. (Wien)* **167** 94

[8] Cewe P, Vorbau R, Omar A, Elmi-Terander A and Edström E 2022 Radiation distribution in a hybrid operating room, utilizing different X-ray imaging systems: investigations to minimize occupational exposure *J. Neurointerv. Surg.* **14** 1139–44

[9] Charles Y P, Al Ansari R, Collinet A, De Marini P, Schwartz J, Nachabe R, Schäfer D, Brendel B, Gangi A and Cazzato R L 2022 Accuracy Assessment of Percutaneous Pedicle Screw Placement Using Cone Beam Computed Tomography with Metal Artifact Reduction *Sensors (Basel)* **22** 4615

[10] Chen H, Zhang Y, Chen Y, Zhang J, Zhang W, Sun H, Lv Y, Liao P, Zhou J and Wang G 2018 LEARN: Learned Experts' Assessment-Based Reconstruction Network for Sparse-Data CT *IEEE Trans. Med. Imaging* **37** 1333–47

[11] Cheng H, Clymer J W, Po-Han Chen B, Sadeghirad B, Ferko N C, Cameron C G and Hinoul P 2018 Prolonged operative duration is associated with complications: a systematic review and meta-analysis *J. Surg. Res.* **229** 134–44

[12] Cleary K and Peters T M 2010 Image-Guided Interventions: Technology Review and Clinical Applications *Annu. Rev. Biomed. Eng.* **12** 119–42

[13] Cormack A M 1963 Representation of a Function by Its Line Integrals, with Some Radiological Applications *J. Appl. Phys.* **34** 2722–7

[14] Cuddy K, Dierks E J, Cheng A, Patel A, Amundson M and Bell R B 2021 Management of Zygomaticomaxillary Complex Fractures Utilizing Intraoperative 3-Dimensional Imaging: The ZYGOMAS Protocol *J. Oral Maxillofac. Surg.* **79** 177–82

[15] Daly M J, Siewerdsen J H, Moseley D J, Jaffray D A and Irish J C 2006 Intraoperative cone-beam CT for guidance of head and neck surgery: Assessment of dose and image quality using a C-arm prototype *Med. Phys.* **33** 3767–80

[16] Dea N, Fisher C G, Batke J, Strelzow J, Mendelsohn D, Paquette S J, Kwon B K, Boyd M D, Dvorak M F S and Street J T 2016 Economic evaluation comparing intraoperative cone beam CT-based navigation and conventional fluoroscopy for the placement of spinal pedicle screws: a patient-level data cost-effectiveness analysis *Spine J.* **16** 23–31

[17] Defrise M and Gullberg G T 2006 Image reconstruction *Phys. Med. Biol.* **51** R139–54

[18] Feldkamp L A, Davis L C and Kress J W 1984 Practical cone-beam algorithm *J. Opt. Soc. Am. A* **1** 612–9

[19] Floridi C, Radaelli A, Abi-Jaoudeh N, Grass M, De Lin M, Chiaradia M, Geschwind J-F, Kobeiter H, Squillaci E, Maleux G, Giovagnoni A, Brunese L, Wood B, Carrafiello G and Rotondo A 2014 C-arm cone-beam computed tomography in interventional oncology: technical aspects and clinical applications *Radiol. Med.* **119** 521–32

[20] Gang G J, Stayman J W, Ouadah S, Ehtiati T and Siewerdsen J H 2015 Task-driven imaging in cone-beam computed tomography *Proc. SPIE Int. Soc. Opt. Eng.* **9412** 941220

[21] Gang G J, Stayman J W, Zbijewski W and Siewerdsen J H 2014 Task-based detectability in CT image reconstruction by filtered backprojection and penalized likelihood estimation *Med. Phys.* **41** 081902

[22] Gelalis I D, Paschos N K, Pakos E E, Politis A N, Arnaoutoglou C M, Karageorgos A C, Ploumis A and Xenakis T A 2012 Accuracy of pedicle screw placement: a systematic review of prospective in vivo studies comparing free hand, fluoroscopy guidance and navigation techniques *Eur. Spine J.* **21** 247–55

[23] Grangeat P 1991 Mathematical framework of cone beam 3D reconstruction via the first derivative of the radon transform *Mathematical Methods in Tomography* (Springer) pp 66–97
[24] Grass M, Koppe R, Klotz E, Proksa R, Kuhn M H, Aerts H, Op de Beek J and Kemkers R 1999 Three-dimensional reconstruction of high contrast objects using C-arm image intensifier projection data *Comput. Med. Imaging Graph.* **23** 311–21
[25] Haines M, Karunaratne Y G, Alani H, Ngo Q, Harish V and Langbart M 2024 Intraoperative computed tomography for the management of traumatic complex facial fractures: a narrative review *Australas. J. Plast. Surg.* **7** 1–5
[26] Hatamikia S, Biguri A, Kronreif G, Figl M, Russ T, Kettenbach J, Buschmann M and Birkfellner W 2021 Toward on-the-fly trajectory optimization for C-arm CBCT under strong kinematic constraints *PLoS One* **16** e0245508
[27] Hisano H, Sakuragi T, Kakiuchi Y, Takita S, Takeda Y, Morita S and Irie H 2017 Usefulness of Intraoperative Cone Beam CT Images on Video-assisted Thoracic Surgery *Nihon Hoshasen Gijutsu Gakkai Zasshi* **73** 87–95
[28] van Hout W M M T, de Kort W W B, Ten Harkel T C, Van Cann E M and Rosenberg A J W P 2022 Zygomaticomaxillary complex fracture repair with intraoperative CBCT imaging. A prospective cohort study *J. Craniomaxillofac. Surg.* **50** 54–60
[29] Hsieh C-P, Hsieh M-J, Fang H-Y and Chao Y-K 2017 Imaging-guided thoracoscopic resection of a ground-glass opacity lesion in a hybrid operating room equipped with a robotic C-arm CT system *J. Thorac. Dis.* **9** E416–9
[30] Hulsen T 2024 Applications of the metaverse in medicine and healthcare *Adv. Lab. Med.* **5** 159–65
[31] International Commission on Radiological Protection 2007 The 2007 Recommendations of the International Commission on Radiological Protection: ICRP Publication 103 *Ann. ICRP* **37** 1–332
[32] Jaju P P and Jaju S P 2015 Cone-beam computed tomography: Time to move from ALARA to ALADA *Imaging Sci. Dent.* **45** 263
[33] Kaiho T, Suzuki H, Hata A, Ito T, Tanaka K, Sakairi Y, Kato H, Shiko Y, Kawasaki Y and Yoshino I 2022 Efficacy and safety of intraoperative cone-beam CT-guided localization of small pulmonary nodules *Interact. Cardiovasc. Thorac. Surg.* **35** ivac236
[34] Kak A C and Slaney M 1988 *Principles of Computerized Tomographic Imaging* (IEEE Press)
[35] Katsevich A 2002a Analysis of an exact inversion algorithm for spiral cone-beam CT *Phys. Med. Biol.* **47** 2583
[36] Katsevich A 2002b Theoretically Exact Filtered Backprojection-Type Inversion Algorithm for Spiral CT *SIAM J. Appl. Math.* **62** 2012–26
[37] Kawasaki T, Kikuchi T, Otani K, Mitsuno Y, Yamao Y, Sawamoto N, Takahashi R and Miyamoto S 2022 Intraoperative cone-beam CT with metal artifact reduction for assessment of the electrode position and the intracranial structures during deep brain stimulation procedure *Acta Neurochir. (Wien)* **164** 2309–16
[38] Kemp P, Stralen J V, Graaf P D, Berkhout E, Horssen P V and Merkus P 2020 Cone-Beam CT Compared to Multi-Slice CT for the Diagnostic Analysis of Conductive Hearing Loss: A Feasibility Study *J. Int. Adv. Otol.* **16** 222–6
[39] Khoury A, Siewerdsen J H, Whyne C M, Daly M J, Kreder H J, Moseley D J and Jaffray D A 2007 Intraoperative cone-beam CT for image-guided tibial plateau fracture reduction *Comput. Aided Surg.* **12** 195–207
[40] Kudo H and Saito T 1990 Feasible cone beam scanning methods for exact reconstruction in three-dimensional tomography *J. Opt. Soc. Am. A* **7** 2169–83
[41] Lahiri A, Maliakal G, Klasky M L, Fessler J A and Ravishankar S 2023 Sparse-View Cone Beam CT Reconstruction Using Data-Consistent Supervised and Adversarial Learning From Scarce Training Data *IEEE Trans. Comput. Imaging* **9** 13–28
[42] Lee S, Gallia G L, Reh D D, Schafer S, Uneri A, Mirota D J, Nithiananthan S, Otake Y, Stayman J W and Zbijewski W 2012 Intraoperative C-arm cone-beam computed tomography: quantitative analysis of surgical performance in skull base surgery. *Laryngoscope* **122** 1925–32
[43] Lee Y C and Lee R 2020 Image-guided pedicle screws using intraoperative cone-beam CT and navigation. A cost-effectiveness study *J. Clin. Neurosci.* **72** 68–71
[44] Markelj P, Tomaževič D, Likar B and Pernuš F 2012 A review of 3D/2D registration methods for image-guided interventions *Med. Image Anal.* **16** 642–61
[45] Mendelsohn D, Strelzow J, Dea N, Ford N L, Batke J, Pennington A, Yang K, Ailon T, Boyd M, Dvorak M, Kwon B, Paquette S, Fisher C and Street J 2016 Patient and surgeon radiation exposure during spinal instrumentation using intraoperative computed tomography-based navigation *Spine J.* **16** 343–54
[46] Mihailidis D N, Stratis A, Gingold E, Carlson R, DeForest W, Gray J, Lally M T, Pizzutiello R, Rong J, Spelic D, Hilohi M C and Massoth R 2024 AAPM Task Group Report 261: Comprehensive quality control methodology and management of dental and maxillofacial cone beam computed tomography (CBCT) systems *Med. Phys.* **51** 3134–64
[47] Muhanna N, Douglas C M, Daly M J, Chan H H L, Weersink R, Townson J, Monteiro E, Yu E, Weimer E, Kucharczyk W, Jaffray D A, Irish J C and de Almeida J R 2021 Evaluating an Image-Guided Operating Room with Cone Beam CT for Skull Base Surgery *J. Neurol. Surg. B Skull Base* **82** e306–14
[48] Noo F, Defrise M, Clackdoyle R and Kudo H 2002 Image reconstruction from fan-beam projections on less than a short scan *Phys. Med. Biol.* **47** 2525–46
[49] Oenning A C, Jacobs R, Salmon B, and DIMITRA Research Group 2021 ALADAIP, beyond ALARA and towards personalized optimization for paediatric cone-beam CT *Int. J. Paediatr. Dent.* **31** 676–8
[50] Pan X, Sidky E Y and Vannier M 2009 Why do commercial CT scanners still employ traditional, filtered back-projection for image reconstruction? *Inverse Probl.* **25** 123009

[51] Parker D L 1982 Optimal short scan convolution reconstruction for fan beam CT *Med. Phys.* **9** 254–7
[52] Pauwels R, Beinsberger J, Collaert B, Theodorakou C, Rogers J, Walker A, Cockmartin L, Bosmans H, Jacobs R, Bogaerts R and Horner K 2012 Effective dose range for dental cone beam computed tomography scanners *Eur. J. Radiol.* **81** 267–71
[53] Pietzka S, Grieser A, Winter K, Schramm A, Metzger M, Semper-Hogg W, Grunert M, Ebeling M, Sakkas A and Wilde F 2024 Comparison of the Effective Radiation Dose in the Region of the Facial Skull Between Multidetector CT, Dental Conebeam CT and Intraoperative 3D C-Arms *Craniomaxillofac. Trauma Reconstr.* **17** 270–8
[54] Pohanka Š, Šimek J, Dolina V, Hozová M and Novotný P 2025 Intraoperative cone beam computed tomography in craniomaxillofacial traumatology – a case report *Acta Chir. Plast.* **67** 252–9
[55] Pohlenz P, Blessmann M, Blake F, Gbara A, Schmelzle R and Heiland M 2008 Major Mandibular Surgical Procedures as an Indication for Intraoperative Imaging *J. Oral Maxillofac. Surg.* **66** 324–9
[56] Pritchett M A, Schampaert S, de Groot J A H, Schirmer C C and van der Bom I 2018 Cone-Beam CT With Augmented Fluoroscopy Combined With Electromagnetic Navigation Bronchoscopy for Biopsy of Pulmonary Nodules *J. Bronchol. Interv. Pulmonol.* **25** 274–82
[57] Rafferty M A, Siewerdsen J H, Chan Y, Daly M J, Moseley D J, Jaffray D A and Irish J C 2006 Intraoperative cone-beam CT for guidance of temporal bone surgery *Otolaryngol. Head Neck Surg.* **134** 801–8
[58] Ramachandran G N and Lakshminarayanan A V 1971 Three-Dimensional Reconstruction from Radiographs and Electron Micrographs: Application of Convolutions Instead of Fourier Transforms *Proc. Natl. Acad. Sci. U.S.A.* **68** 2236–40
[59] Samei E, Bakalyar D, Boedeker K L, Brady S, Fan J, Leng S, Myers K J, Popescu L M, Ramirez Giraldo J C, Ranallo F, Solomon J, Vaishnav J and Wang J 2019 Performance evaluation of computed tomography systems: Summary of AAPM Task Group 233 *Med. Phys.* **46** e735–56
[60] Scarfe W C, Farman A G and Sukovic P 2006 Clinical applications of cone-beam computed tomography in dental practice *J. Can. Dent. Assoc.* **72** 75–80
[61] Schafer S, Nithiananthan S, Mirota D J, Uneri A, Stayman J W, Zbijewski W, Schmidgunst C, Kleinszig G, Khanna A J and Siewerdsen J H 2011 Mobile C-arm cone-beam CT for guidance of spine surgery: Image quality, radiation dose, and integration with interventional guidance *Med. Phys.* **38** 4563–74
[62] Schulze R, Heil U, Groβ D, Bruellmann D, Dranischnikow E, Schwanecke U and Schoemer E 2011 Artefacts in CBCT: a review *Dentomaxillofac. Radiol.* **40** 265–73
[63] Setser R, Chintalapani G, Bhadra K and Casal R F 2020 Cone beam CT imaging for bronchoscopy: a technical review *J. Thorac. Dis.* **12** 7416–28
[64] Shao L, Tang W, Zhang Z and Chen X 2023 Medical metaverse: technologies, applications, challenges and future *J. Mech. Med. Biol.* **23** 2350028
[65] Sharma M and Deogaonkar M 2016 Accuracy and safety of targeting using intraoperative "O-arm" during placement of deep brain stimulation electrodes without electrophysiological recordings *J. Clin. Neurosci.* **27** 80–6
[66] Shepp L A and Logan B F 1974 The Fourier reconstruction of a head section *IEEE Trans. Nucl. Sci.* **21** 21–43
[67] Siewerdsen J and Stayman, J. W. 2013 Task-Based Trajectories in Iteratively Reconstructed Interventional Cone-Beam CT *Proceedings of the 12th International Meeting on Fully Three-Dimensional Image Reconstruction in Radiology and Nuclear Medicine* pp 257–60
[68] Smith B D 1985 Image reconstruction from cone-beam projections: necessary and sufficient conditions and reconstruction methods *IEEE Trans. Med. Imaging* **4** 14–25
[69] Sonke J-J, Zijp L, Remeijer P and van Herk M 2005 Respiratory correlated cone beam CT *Med. Phys.* **32** 1176–86
[70] Spin-Neto R and Wenzel A 2016 Patient movement and motion artefacts in cone beam computed tomography of the dentomaxillofacial region: a systematic literature review *Oral Surg. Oral Med. Oral Pathol. Oral Radiol.* **121** 425–33
[71] Stock M, Pasler M, Birkfellner W, Homolka P, Poetter R and Georg D 2009 Image quality and stability of image-guided radiotherapy (IGRT) devices: A comparative study *Radiother. Oncol.* **93** 1–7
[72] Supanich M P, Siewerdsen J H, Fahrig R, Farahani K, Gang G J, Helm P, Jans J, Jones K, Koenig T, Kuhls-Gilcrist A T, Lin M, Riddell C, Ritschl L, Schafer S, Schueler B A, Silver M D, Timmer J, Trousset Y and Zhang J 2023 AAPM Task Group Report 238: 3D C-arms with volumetric imaging capability *Med. Phys.* **50** e904–45
[73] Thomas C, Dong G, Schonebaum D I, Challa S, Adams A J, Song E, Arif F, Foppiani J A, Schubert W, Choudry U and Lin S J 2026 The Current State of Intraoperative Imaging in Maxillofacial Surgery: A Systematic Review *J. Clin. Med.* **15** 1675
[74] Toroi P, Kaasalainen T, Uusi-Simola J, Aho P, Mäkelä T and Kortesniemi M 2023 Intraoperative CBCT imaging in endovascular abdomen aneurysm repair - Optimization of exposure parameters using a stent phantom *Phys. Med.* **112** 102634
[75] Tuy H K 1983 An Inversion Formula for Cone-Beam Reconstruction *SIAM J. Appl. Math.* **43** 546–52
[76] Unger S, Penzenstadler M, Al-Haj Husain A, Wiedemeier D, Stadlinger B and Valdec S 2023 Comparison of Geometric Accuracy of Low-Dose and Standard-Dose Dental CBCT Imaging Protocols in CAD/CAM-Guided Dental Implant Surgery *Int. J. Oral Maxillofac. Implants* **38** 287–94
[77] Verdun F R, Racine D, Ott J G, Tapiovaara M J, Toroi P, Bochud F O, Veldkamp W J H, Schegerer A, Bouwman R W, Giron I H, Marshall N W and Edyvean S 2015 Image quality in CT: From physical measurements to model observers *Phys. Med.* **31** 823–43

[78] Vorbau R, Hulthén M and Omar A 2024 Task-based image quality assessment of an intraoperative CBCT for spine surgery compared with conventional CT *Phys. Med.* **124** 103426
[79] Wang A S, Stayman J W, Otake Y, Kleinszig G, Vogt S, Gallia G L, Khanna A J and Siewerdsen J H 2014 Soft-tissue imaging with C-arm cone-beam CT using statistical reconstruction *Phys. Med. Biol.* **59** 1005–26
[80] Wang G, Ye J C, Mueller K and Fessler J A 2018 Image Reconstruction is a New Frontier of Machine Learning *IEEE Trans. Med. Imaging* **37** 1289–96
[81] Weinstein J N, Spratt K F, Spengler D, Brick C and Reid S 1988 Spinal pedicle fixation: reliability and validity of roentgenogram-based assessment and surgical factors on successful screw placement *Spine (Phila Pa 1976)* **13** 1012–8
[82] Wu P, Tersol A, Clackdoyle R, Boone J M and Siewerdsen J H 2023 Cone-beam CT sampling incompleteness: analytical and empirical studies of emerging systems and source-detector orbits *J. Med. Imaging (Bellingham)* **10** 033503
[83] Yang S-M, Ko W-C, Lin M-W, Hsu H-H, Chan C-Y, Wu I-H, Chang Y-C and Chen J-S 2016 Image-guided thoracoscopic surgery with dye localization in a hybrid operating room *J. Thorac. Dis.* **8** S681–9
[84] Zellerhoff M, Scholz B, Ruehrnschopf E-P and Brunner T 2005 Low contrast 3D reconstruction from C-arm data *Proc. SPIE* vol 5745 pp 646–55
[85] Zhang X, Sisniega A, Zbijewski W B, Lee J, Jones C K, Wu P, Han R, Uneri A, Vagdargi P, Helm P A, Luciano M, Anderson W S and Siewerdsen J H 2023 Combining physics-based models with deep learning image synthesis and uncertainty in intraoperative cone-beam CT of the brain *Med. Phys.* **50** 2607–24